\documentclass{article}

\usepackage{microtype}
\usepackage{graphicx}
\usepackage{subfigure}
\usepackage{booktabs} 

\usepackage{hyperref}

\usepackage[accepted]{icml2025}

\usepackage{amsmath}
\usepackage{amssymb}
\usepackage{mathtools}
\usepackage{amsthm}

\usepackage[capitalize,noabbrev]{cleveref}

\theoremstyle{plain}

\theoremstyle{definition}

\theoremstyle{remark}

\usepackage[textsize=tiny]{todonotes}

\icmltitlerunning{New Interaction Paradigm for Complex EDA Software Leveraging GPT}

\begin{document}

\twocolumn[
\icmltitle{New Interaction Paradigm for Complex EDA Software Leveraging GPT}



\begin{icmlauthorlist}
\icmlauthor{Xinyu Wang*}{mcgill,simpleway}
\icmlauthor{Boyu Han*}{stanford,simpleway}
\icmlauthor{Zhenghan Tai}{toronto,simpleway}
\icmlauthor{Jingrui Tian}{ucla}
\icmlauthor{Yifan Wang}{tsinghua,smarton}
\icmlauthor{Junyu Yan}{buaa,smarton}
\icmlauthor{Yidong Tian}{manchester,smarton}

\end{icmlauthorlist}

\icmlaffiliation{mcgill}{McGill University, Canada}
\icmlaffiliation{stanford}{Stanford University, USA}
\icmlaffiliation{toronto}{University of Toronto, Canada}
\icmlaffiliation{tsinghua}{Tsinghua University, China}
\icmlaffiliation{buaa}{Beihang University, China}
\icmlaffiliation{manchester}{The University of Manchester, UK}
\icmlaffiliation{ucla}{University of California, Los Angeles, USA}
\icmlaffiliation{smarton}{Beijing Smarton Empower Co., Ltd., China}
\icmlaffiliation{simpleway}{SimpleWay AI, China}

\icmlcorrespondingauthor{Xinyu Wang}{xinyuwang@mail.mcgill.ca}
\icmlcorrespondingauthor{Boyu Han}{boyuhan@stanford.edu}

\icmlkeywords{Machine Learning, ICML}

\vskip 0.3in
]



\printAffiliationsAndNotice{\icmlEqualContribution} 

\begin{abstract}
Electronic Design Automation (EDA) tools such as KiCad offer powerful functionalities but remain difficult to use, particularly for beginners, due to their steep learning curves and fragmented documentation. To address this challenge, we present SmartonAI, an AI-assisted interaction system that integrates large language models into the EDA workflow, enabling natural language communication, intelligent task decomposition, and contextual plugin execution. SmartonAI consists of two main components: a Chat Plugin that breaks down user instructions into subtasks and retrieves tailored documentation, and a OneCommandLine Plugin that recommends and executes relevant plugins based on user intent. The system supports multilingual interaction and adapts to user feedback through incremental learning. Preliminary results suggest that SmartonAI significantly reduces onboarding time and enhances productivity, representing a promising step toward generalizable AI-assisted interaction paradigms for complex software systems.
\end{abstract}
\section{Introduction}

Electronic Design Automation (EDA) tools such as KiCad~\cite{kicad-official}, Cadence~\cite{cadence-official}, and Altium Designer~\cite{altium-official} provide powerful capabilities for schematic capture, PCB layout, and verification. Despite their robustness, these tools remain notoriously difficult to master due to steep learning curves, intricate user interfaces, and a lack of intuitive, task-driven guidance. For novice designers in particular, identifying the right features or plugins often requires sifting through fragmented documentation or community forums, which severely limits productivity and accessibility. Furthermore, these resources are often static, outdated, or written for expert audiences, offering little assistance when users encounter nuanced, context-specific challenges during design iterations. The lack of an adaptive, interactive support mechanism means that users must engage in time-consuming trial-and-error, leading to frustration and reduced design throughput.

Recent advances in large language models (LLMs) have transformed how users engage with complex software systems. Architecturally, LLMs are typically based on the Transformer framework~\cite{vaswani2017attention}, which enables attention-based modeling of long-range dependencies in text. Pretrained on large-scale corpora and fine-tuned for downstream tasks, modern LLMs such as GPT-4.1~\cite{gpt4-openai}, Claude 3 Opus~\cite{claude3-anthropic}, Gemini 2.5 Pro~\cite{gemini25-deepmind}, and open-source systems like Meta's LLaMA-3~\cite{llama3-grattafiori2024} and Alibaba's Qwen2.5~\cite{qwen25-alibaba} exhibit remarkable capabilities in few-shot learning, tool use, and contextual reasoning. These models maintain conversational state across multiple turns, support function calling, and exhibit emergent capabilities such as code generation, retrieval augmentation, and planning. As a result, they are increasingly being integrated into real-world workflows through APIs, plugins, and autonomous agents.

Complementing these advances, retrieval-augmented generation (RAG) techniques~\cite{rag-lewis2020} have become essential in improving LLM effectiveness in knowledge-intensive tasks. RAG pipelines dynamically retrieve relevant external content—such as documentation, plugin descriptions, or design tutorials—based on user queries, and condition the language model's output on this evidence. This hybrid framework helps mitigate context window limitations and ensures generated responses are grounded in factual, domain-specific knowledge. In the context of EDA software, such augmentation is particularly beneficial, as it enables accurate and context-aware guidance drawn directly from official manuals or plugin repositories.

LLM-based interfaces have demonstrated success in domains such as software development (e.g., Codex~\cite{codex-chen2021}), general AI orchestration (e.g., HuggingGPT~\cite{hugginggpt-shen2023}), and multimodal agents. However, their application in domain-specific engineering tools—such as EDA software—remains underexplored. Unlike general-purpose tasks, EDA workflows involve intricate procedural logic, structured design hierarchies, and tightly-coupled UI actions, which pose unique challenges for general LLM deployment.

In this work, we introduce \textbf{SmartonAI}, an AI-assisted interaction system for EDA software, initially integrated with KiCad~\cite{kicad-official}. SmartonAI leverages the natural language understanding and reasoning capabilities of LLMs to transform vague or high-level user requests into concrete design operations. The system is composed of two complementary components: (1) a \textbf{Chat Plugin}, which supports multi-turn dialogue for decomposing tasks and retrieving relevant documentation, and (2) a \textbf{OneCommandLine Plugin}, which enables intelligent plugin recommendation and parameterized execution based on user goals. Both components integrate retrieval-augmented generation~\cite{rag-lewis2020} to enable precise document grounding and plugin-specific assistance.

SmartonAI pioneers the application of LLMs and RAG pipelines in EDA workflows, bridging the gap between user intent and low-level design execution. By combining language-based guidance with context-aware automation, SmartonAI lowers the barrier to entry for PCB design, improves user efficiency, and offers a blueprint for extending AI-assisted interfaces to other complex, domain-specific software ecosystems.

\section{Related Work}

Our work builds upon and extends multiple lines of research at the intersection of human-AI interaction, large language model (LLM) orchestration, retrieval-augmented generation (RAG), and computer-aided design automation.

\textbf{Natural language interfaces for software engineering.} The emergence of LLM-powered assistants has catalyzed a paradigm shift in software tooling. GitHub Copilot, based on OpenAI Codex and later GPT-4, exemplifies the integration of natural language programming into mainstream development workflows by translating user intent into executable code \cite{Li_2022}. Similarly, tools such as ChatGPT, Copilot for Office, Notion AI, and CodeWhisperer have demonstrated that natural language interfaces can effectively support complex, multi-modal tasks spanning code generation, document editing, and data analysis \cite{Nam2024LLMCodeUnderstanding,Wang2023NL2Code}. These systems rely heavily on few-shot prompting, in-context learning, and tool use APIs. More recent models such as Claude 3 Opus, Gemini 2.5, and open-source models like Qwen2.5 and LLaMA-3 continue to push the boundaries of long-context reasoning, tool calling, and multi-agent collaboration, suggesting even greater potential for domain-specific adaptation.

\textbf{LLM-based task orchestration and autonomous agents.} HuggingGPT \cite{Shen2023HuggingGPT} and similar agentic frameworks such as LangChain, Auto-GPT, and MetaGPT demonstrate how LLMs can coordinate subtasks across a suite of tools and models, effectively functioning as high-level controllers \cite{Schick2023Toolformer,Yao2023ReAct}. These systems emphasize modularity and dynamic reasoning, enabling LLMs to decompose user instructions and interact with APIs, search tools, plug-ins, or even simulators. Several efforts have extended these frameworks to encompass planning, memory, and reflection modules, further enhancing robustness for real-world deployments. However, despite their generalizability, such orchestration pipelines are rarely applied in engineering tools with graphical user interfaces (GUIs), especially where stateful, real-time manipulation of design artifacts is required.

\textbf{Retrieval-Augmented Generation (RAG).} RAG architectures improve LLM performance by retrieving relevant external knowledge to supplement the model’s context window. Notable systems such as REALM \cite{Guu2020REALM}, RAG \cite{Lewis2020RAG}, Atlas \cite{izacard2022atlasfewshotlearningretrieval}, and RETRO \cite{Borgeaud2022RETRO} have demonstrated superior performance on question-answering and knowledge-intensive tasks. These methods integrate neural retrievers with generative models, allowing users to issue natural language queries whose responses are grounded in structured corpora. In addition, recent applications of RAG in software documentation alignment, legal reasoning, and multi-step scientific QA highlight its effectiveness in structured, domain-specific scenarios.

Within the domain of EDA tools, SmartonAI applies RAG techniques through a module named \textit{DocHelper}, which indexes and embeds tool-specific documentation—including official manuals, plugin descriptions, and usage examples. When users pose natural language queries, DocHelper retrieves the most semantically relevant content using dense retrieval, and conditions the LLM response on this retrieved evidence. This allows SmartonAI to deliver grounded, context-aware explanations and actionable suggestions, tailored to the user’s current task and environment. While prior RAG research has mostly focused on open-domain settings, our work demonstrates its value in GUI-driven, real-time engineering workflows, where fine-grained tool usage guidance is critical but hard to encode manually.

\textbf{Intelligent automation in EDA tools.} The EDA community has historically emphasized automation through domain-specific algorithms, with major progress in placement, routing, logic synthesis, and verification. Tools such as DeepPCB, AutoDMP, and DreamPlace have leveraged ML to enhance design quality and efficiency in backend workflows \cite{Mirhoseini2021ChipPlacement,Cheng2022ChipPG,Pei2023AlphaSyn}. Concurrently, efforts in human-in-the-loop EDA remain nascent. GUI scripting, design wizards, and parameterized templates offer partial support for user guidance, but these approaches are often brittle, inflexible, and non-interactive. Furthermore, the majority of prior work overlooks the early-phase design exploration and iterative debugging stages, where human-AI collaboration could have the greatest impact.

\textbf{Positioning of SmartonAI.} In contrast to prior work, SmartonAI introduces a human-centric AI interface tailored for front-end EDA interaction. It is the first, to our knowledge, to tightly integrate multi-turn conversational LLMs with real-time KiCad tool invocation, plugin discovery, RAG-based document retrieval, and adaptive HTML documentation assembly. Unlike generic code generators or rule-based scripting, SmartonAI dynamically interprets user queries, breaks them down into actionable tasks, and orchestrates both explanation and execution flows through a unified natural language interface. This approach not only enhances usability and accessibility for novice users, but also provides experienced designers with a scalable, language-driven control layer for rapid prototyping and design iteration.

\section{Method}

\subsection{System Overview}

SmartonAI is designed as a modular AI-assistant system for EDA tools, integrating large language models (LLMs), retrieval-based document grounding, and plugin execution interfaces. It comprises two main components: the Chat Plugin and the OneCommandLine Plugin. Both modules are optimized for task decomposition, contextual guidance, and direct interaction with KiCad through a unified conversational interface.

\begin{figure}[htbp]
    \centering
    \includegraphics[width=0.9\linewidth]{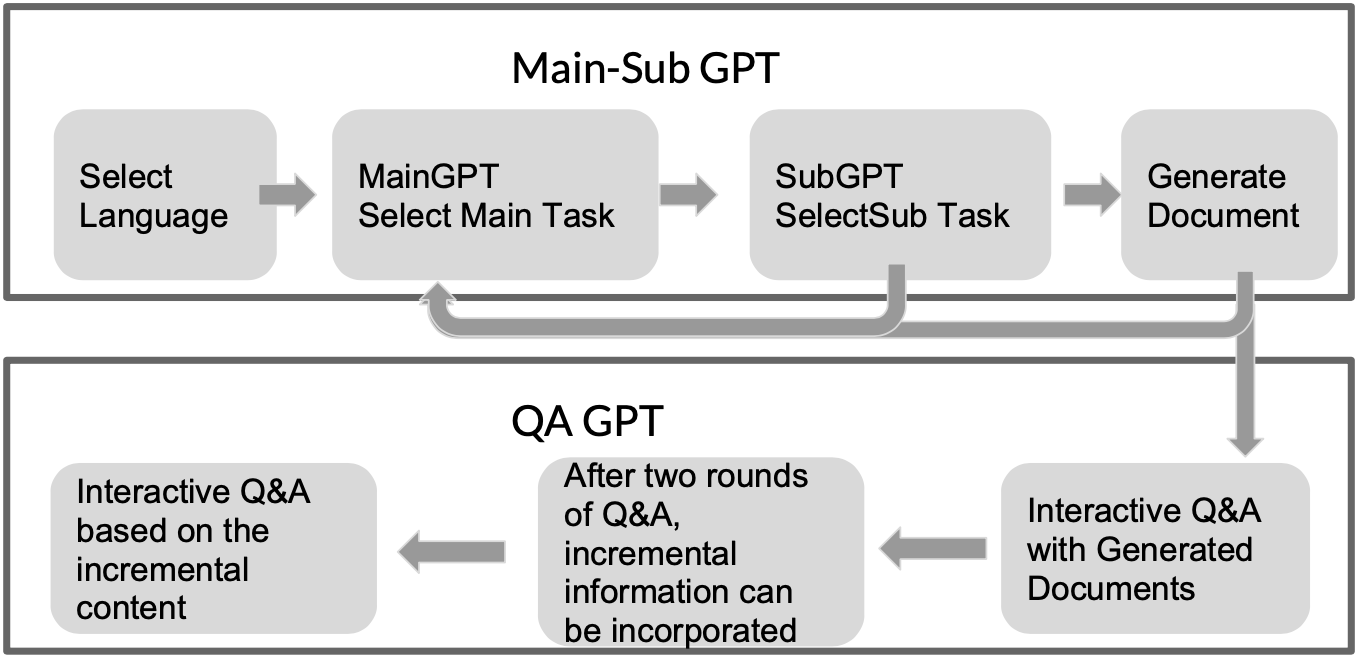}
    \caption{Overview of The Chat Plugin}
    \label{fig:chat_plugin}
\end{figure}

Figure~\ref{fig:chat_plugin} illustrates the architecture of the Chat Plugin. The system begins with language selection and routes user requests through a hierarchical reasoning flow involving a main task selector (MainGPT) and a sub-task selector (SubGPT). The chosen task guides the retrieval and generation of tailored documentation, which is then passed to a QA module. The QA GPT enables iterative dialogue grounded on the generated content, supporting incremental learning and refinement across multiple user turns.

\begin{figure}[htbp]
    \centering
    \includegraphics[width=0.9\linewidth]{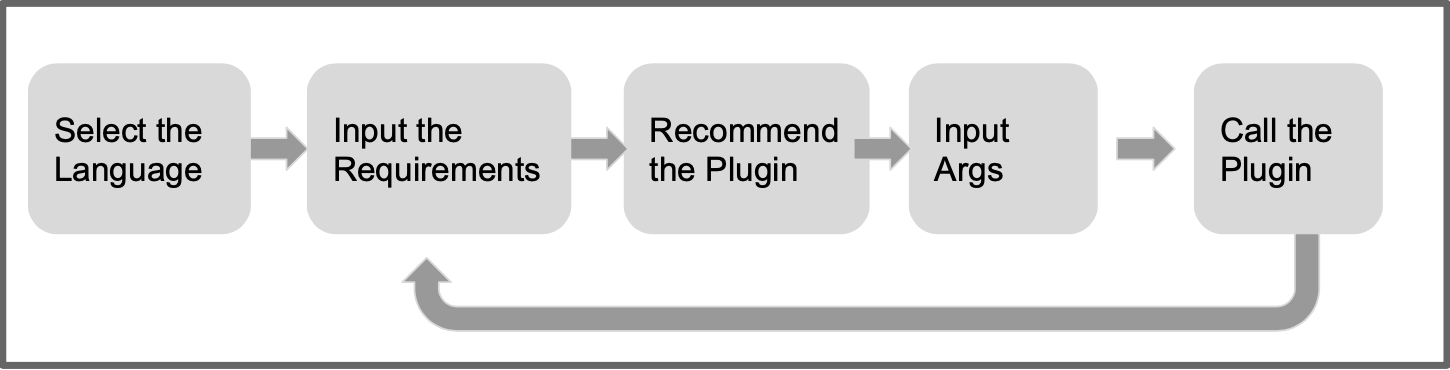}
    \caption{Overview of The OneCommandLine Plugin}
    \label{fig:commandline_plugin}
\end{figure}

Figure~\ref{fig:commandline_plugin} shows the architecture of the OneCommandLine Plugin. Users input requirements in natural language, and the system recommends appropriate KiCad plugins via semantic matching. After gathering parameter inputs, the plugin is executed with minimal manual intervention. The interaction loop supports multilingual input, plugin auto-completion, and feedback-based correction.

These figures provide a conceptual overview of the SmartonAI system and highlight its conversational structure, dynamic decision routing, and seamless integration with domain-specific tooling. The following subsections detail each module’s mechanism.

\subsection{Chat Plugin: Interactive Task Decomposition}
\label{sec:chat_plugin}

The Chat Plugin facilitates context-aware, multi-turn interactions by decomposing vague user intents into concrete, actionable design steps. It comprises two cascaded LLM components: \textbf{Main-Sub GPT} and \textbf{QA GPT}.

The \textbf{Main-Sub GPT} module performs hierarchical task classification and planning. Given a natural language query, the MainGPT model classifies the user’s intent into one of 20 predefined macro-task categories (e.g., ``netlist verification,'' ``footprint adjustment''). This is implemented using instruction-tuned LLMs (e.g., Qwen2.5, LLaMA-3) augmented with custom prompt templates and task ontologies. The predicted macro-task is passed to a SubGPT, which selects one or more domain-specific subtasks via dense retrieval over a curated task database. This modular decomposition enables interpretable routing, few-shot prompt specialization, and plug-and-play extensibility.

The selected subtasks trigger document synthesis routines and context gathering from the DocHelper. These outputs, along with the user query, are formatted into structured prompt templates and passed to the \textbf{QA GPT}, which maintains conversational state across multiple user rounds. The QA GPT answers queries using a mixture of retrieval-based prompting, RAG grounding, and constrained decoding to improve factuality and clarity.

To improve robustness, the plugin supports dynamic feedback injection, where users can mark responses as ``unsatisfactory'' and provide clarifying constraints. These are encoded using structured system prompts that modify retrieval or generation pipelines. In practice, this allows SmartonAI to support both guided learning and exploratory design workflows.

\subsection{DocHelper: Retrieval-Augmented Document Grounding}

The DocHelper subsystem provides retrieval-augmented grounding for task-aware question answering. It maintains an index over segmented documentation sources, including official KiCad manuals, plugin metadata, code examples, and community Q\&A posts.

Documents are preprocessed using a hybrid pipeline: HTML and Markdown files are chunked into overlapping spans with dynamic window sizes based on semantic boundaries. Each chunk is embedded using a transformer-based encoder model (e.g., BGE-M3, E5-large) and stored in a FAISS vector store with metadata tags (e.g., tool version, component type).

During interaction, each subtask predicted by the Chat Plugin issues a query to the DocHelper index. These queries are constructed via a learned retriever-query generator pipeline or statically templated from subtask descriptors. Retrieved chunks are ranked with a hybrid BM25 and dense similarity score, filtered by task type and user context, and assembled into a single context block for prompt injection.

To enable high accuracy and controllability, DocHelper supports:
\begin{itemize}
  \item \textbf{Context distillation}: Concise prompt-level summarization using GPT-4 or compression-augmented LLMs;
  \item \textbf{Source attribution}: Inline citation or reference links to raw HTML sources;
  \item \textbf{Incremental retrieval}: Feedback-based query rewriting to overcome failures in initial retrieval.
\end{itemize}

Overall, the DocHelper enables SmartonAI to deliver grounded, context-sensitive answers that are both accurate and traceable—critical in domains like EDA where users rely on tool-specific semantics and fine-grained usage details.

\subsection{OneCommandLine Plugin: Plugin Recommendation and Execution}

The OneCommandLine Plugin enables zero-shot plugin invocation from natural language by integrating semantic parsing, plugin metadata grounding, and parameter validation into a coherent workflow. It is particularly suited for users who prefer action-oriented interactions without navigating nested menus or documentation.

The workflow begins by parsing the user input into structured intents using an LLM-based semantic parser. This parser maps the user's natural language request to a latent representation of task goals and expected arguments, leveraging a combination of slot-filling techniques and prompt-conditioned generation (e.g., using in-context few-shot examples).

Next, a plugin retriever ranks available KiCad plugins using dense-sparse hybrid retrieval. Plugin metadata—consisting of descriptions, function signatures, input constraints, and usage examples—is indexed offline using SBERT and BM25. At runtime, a multi-vector MaxSim scoring function is used to match user intents against the plugin corpus.

Upon selecting the most relevant plugin, the system dynamically generates parameter input forms or argument templates. These templates include auto-suggested values based on plugin schema definitions and previous user sessions, enhancing usability and reducing invalid configurations.

Execution is handled by a backend KiCad bridge layer, which abstracts plugin calling into JSON-RPC requests. This layer verifies argument types, applies necessary data normalization, and invokes the plugin in the native environment. Any returned results or error traces are passed back to the LLM layer for summarization and user display.

The OneCommandLine Plugin further supports:
\begin{itemize}
  \item \textbf{Interactive clarification}: When ambiguous inputs are detected, the system engages the user with disambiguation prompts;
  \item \textbf{Multilingual compatibility}: Prompts and plugin metadata are translated using a multilingual embedding model (e.g., LaBSE);
  \item \textbf{User feedback loop}: Plugin invocation results are logged and reused to fine-tune recommendation heuristics.
\end{itemize}

Together, this module operationalizes a command-driven interface for non-expert users and enhances the productivity of experienced engineers by minimizing context switches and manual plugin lookups.

\subsection{Implementation Details}

The SmartonAI system is implemented as a modular, production-grade application that integrates LLM inference, document retrieval, plugin execution, and frontend rendering in a unified architecture.

\paragraph{Frontend Infrastructure.} The desktop interface is built using PyQt5, providing a native, responsive GUI that supports multilingual input, documentation rendering, and real-time chat display. To enable fluid user experience and minimize response latency, the system leverages Server-Sent Events (SSE) for streaming LLM responses, allowing partial completions to be shown token-by-token in the chat window. This improves interactivity during long-form responses and enables early interruption or correction. The frontend also includes dynamically generated form components for plugin execution, parameter filling, and trace display.

\paragraph{Backend Runtime.} The backend consists of three service layers:
\begin{itemize}
    \item \textbf{LLM Layer:} We employ Qwen2.5-0.5B and LLaMA-3-8B as our core models, deployed via vLLM for high-throughput, low-latency serving. Instruction tuning is applied using LoRA adapters fine-tuned on domain-specific EDA queries. Generation is managed via a sliding context window with ChatML-style prompts. In addition to local model inference, we support cloud-based APIs including OpenAI's GPT-4.0 and Gemini 1.5 Pro, which are used for fallback, comparative evaluation, and specific tasks such as summarization or zero-shot classification. For open-source models like Mistral and Deepseek, we utilize Ollama as a lightweight containerized serving layer, enabling model swapping without modifying the orchestration code.

    \item \textbf{Retrieval Layer:} The DocHelper subsystem uses FAISS (HNSW index) to serve vector-based search queries over HTML-parsed KiCad documentation. Embeddings are computed using BGE-M3 or E5-large models and augmented with sparse signals (BM25) for hybrid ranking. When documents include images (e.g., annotated schematics, UI screenshots), we extract the images and compute vision embeddings using a CLIP-like encoder. These embeddings are linked to the corresponding text chunks, allowing the retriever to match queries to visual content. This multimodal indexing improves the ability to answer questions involving visual UI layout, button locations, or design schematics.

    \item \textbf{Plugin Execution Layer:} A JSON-RPC bridge abstracts access to KiCad's internal Python API. Plugins are described in YAML-based schemas, which drive auto-generated UI components and parameter validation. The bridge is robust to runtime exceptions and provides structured feedback for LLM summarization.
\end{itemize}

\paragraph{Asynchronous Orchestration.} The backend services are connected using asynchronous FastAPI routes, with SSE channels for token streaming and WebSocket fallback. Tasks such as plugin recommendation, document retrieval, and chat response are coordinated using asyncio event loops to ensure non-blocking I/O and modular debugging.

\paragraph{Monitoring and Logging.} All user interactions are logged for quality improvement and debugging. Logs include LLM prompts, retrieval hits, selected plugin metadata, execution success/failure, and user feedback ratings. These logs are used for downstream fine-tuning, error tracing, and evaluation benchmarking.

This architecture ensures that SmartonAI is not only effective in guiding users through complex EDA workflows but also robust, extensible, and suitable for continuous deployment in production environments.

\section{Experiments}

\subsection{Overview of Evaluation Setup}
We assess SmartonAI within the KiCad environment across a diverse set of representative Electronic Design Automation (EDA) workflows. Experiments are run on a macOS Ventura 13.6 system equipped with a 16-core Apple M1 Pro CPU and 32 GB of RAM. Inference is powered by a hybrid backend using vLLM to support Qwen2.5-0.5B for efficient, memory-optimized decoding, and LLaMA3-8B for enhanced reasoning capacity in more complex tasks. 

Our retrieval stack leverages DocHelper, which indexes HTML-structured design manuals, plugin documentation, and scripting guides using FAISS with dense vector representations. The retrieval system enables paragraph-level grounding, allowing the language model to cite exact documentation snippets during multi-turn interactions. Evaluation dimensions span natural language understanding (NLU), procedural toolchain reasoning, zero-shot plugin selection, and multi-step planning for command synthesis and execution.

\subsection{Use Case 1: Multi-Turn Task Decomposition in Chat Plugin}
\begin{figure}[ht]
    \centering
    \subfigure[]{\includegraphics[width=0.45\textwidth]{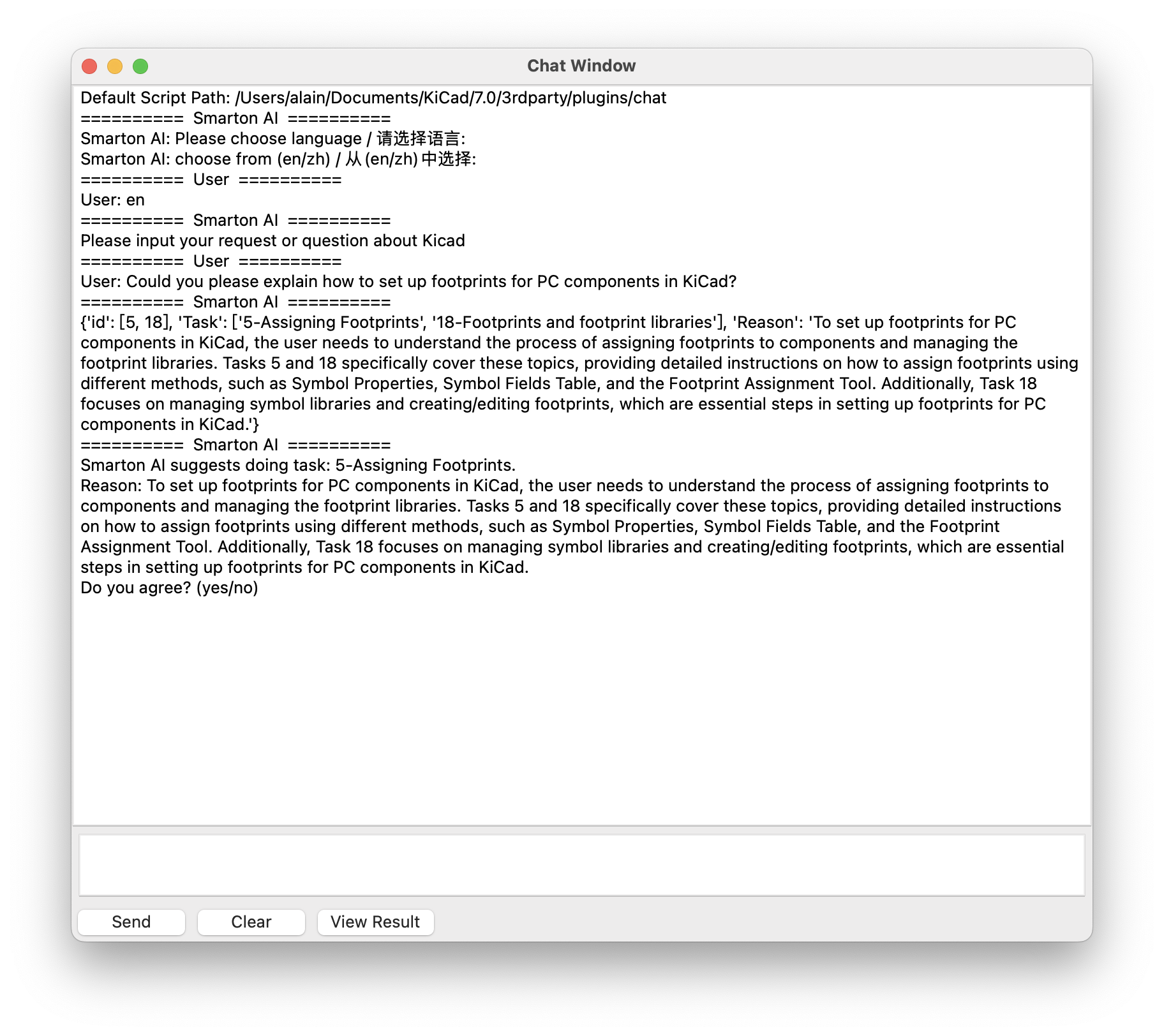}}
    \subfigure[]{\includegraphics[width=0.45\textwidth]{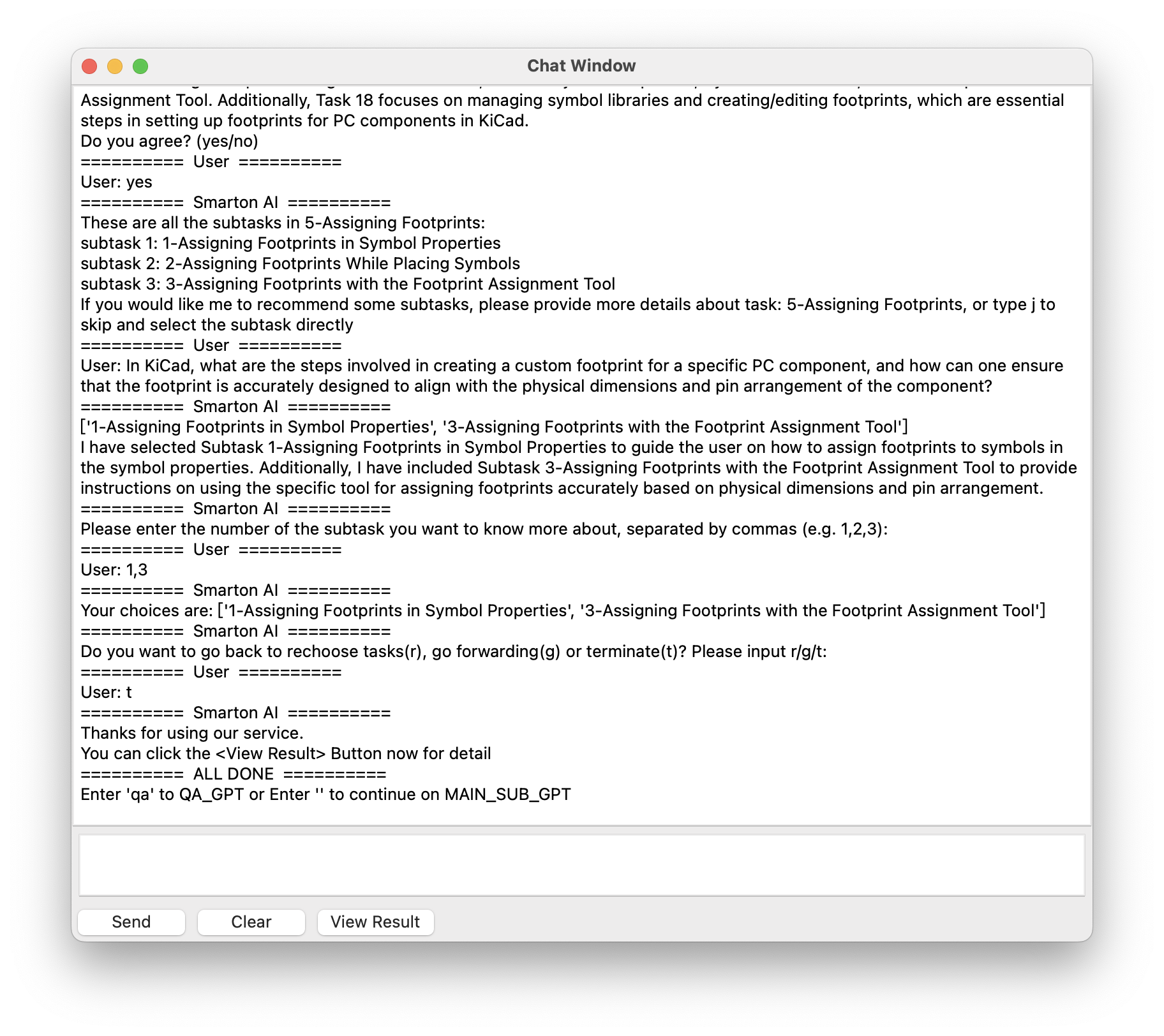}}
    \caption{(a) Selection of Main Tasks, (b) Selection of Sub Tasks}
    \label{fig:chat_plugin_demo}
\end{figure}

Figure~\ref{fig:chat_plugin_demo} illustrates the Chat Plugin’s interactive agent-based planning pipeline. Upon receiving a high-level user query (e.g., “assign footprints to components”), SmartonAI’s MainGPT performs intent classification and workflow segmentation. SubGPT then generates a task graph by decomposing the query into actionable subtasks, which are further refined through user confirmation prompts.

Throughout this interaction, QA-GPT continuously grounds clarifications and generated plans in the retrieved design documentation. The web-based interface dynamically renders the most relevant documentation spans, aiding users in decision-making. This pipeline demonstrates the ability of SmartonAI to maintain dialogue state, validate user goals, and adaptively refine its planning strategy.

\begin{figure}[ht]
    \centering
    \subfigure[]{\includegraphics[width=0.45\textwidth]{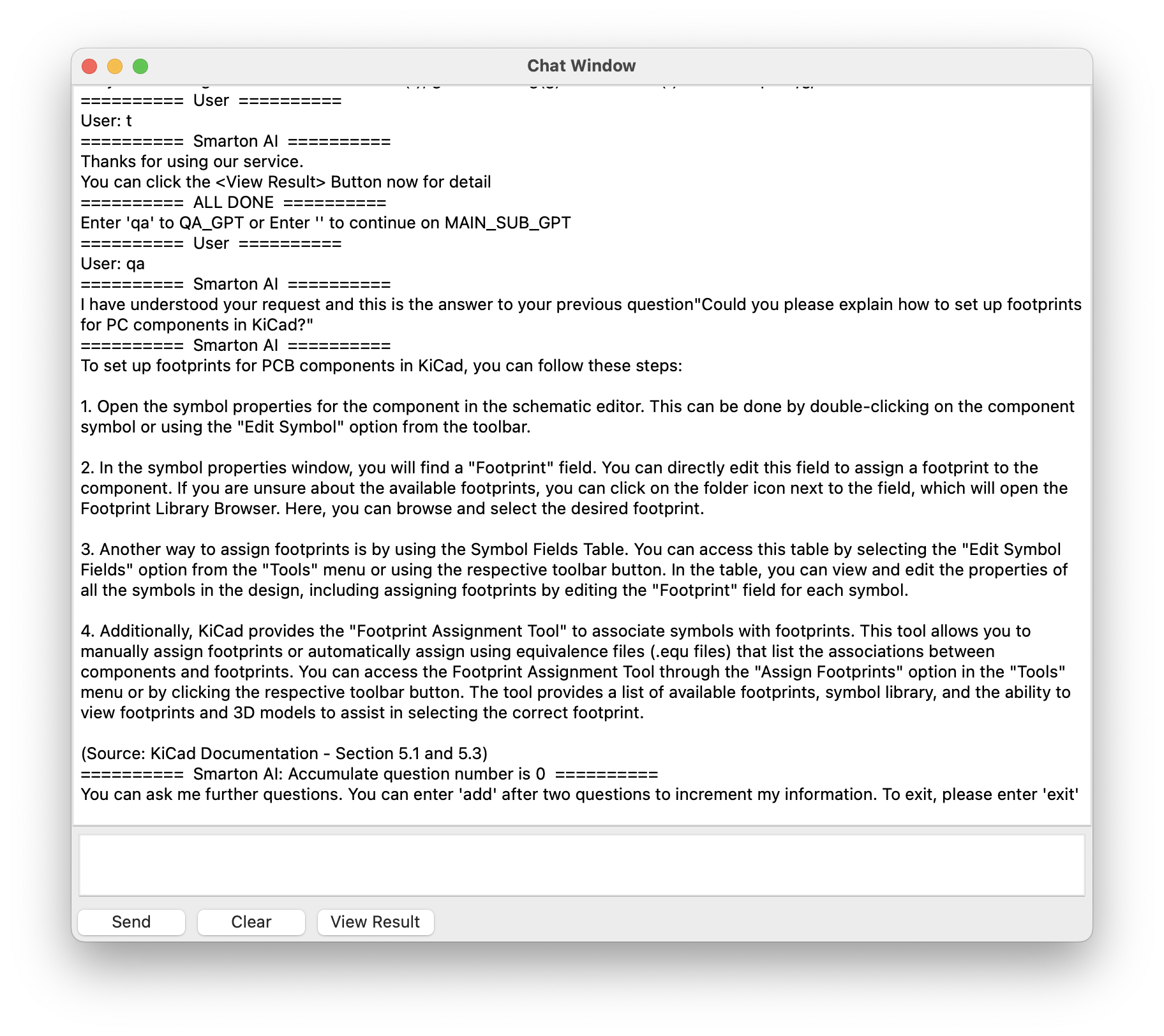}}
    \subfigure[]{\includegraphics[width=0.45\textwidth]{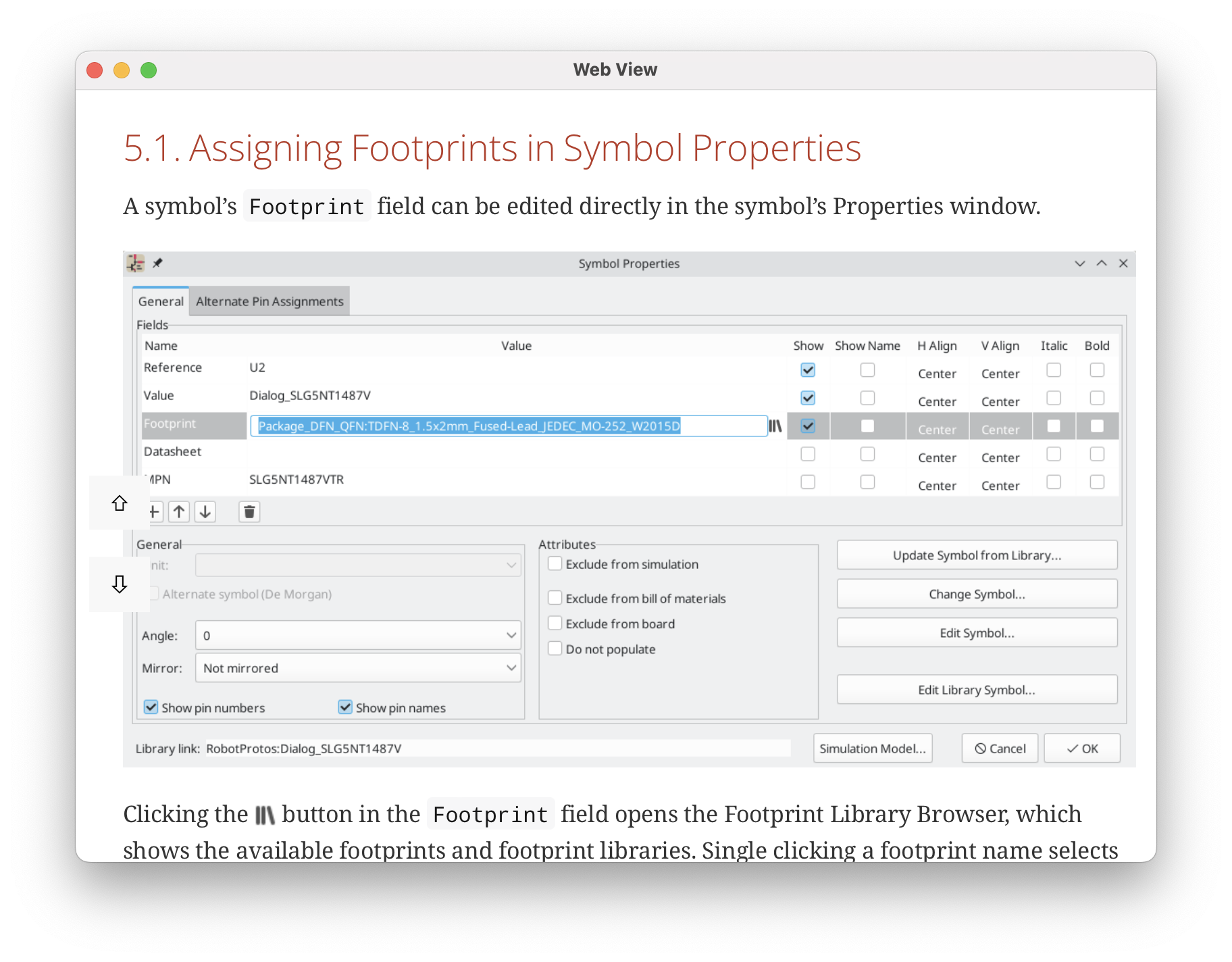}}
    \caption{(c) QA GPT Interaction, (d) Tailored Documentation Rendered in Web View}
\end{figure}

\subsection{Use Case 2: Plugin Recommendation and Execution in OneCommandLine Plugin}
\begin{figure}[ht]
    \centering
    \includegraphics[width=0.48\textwidth]{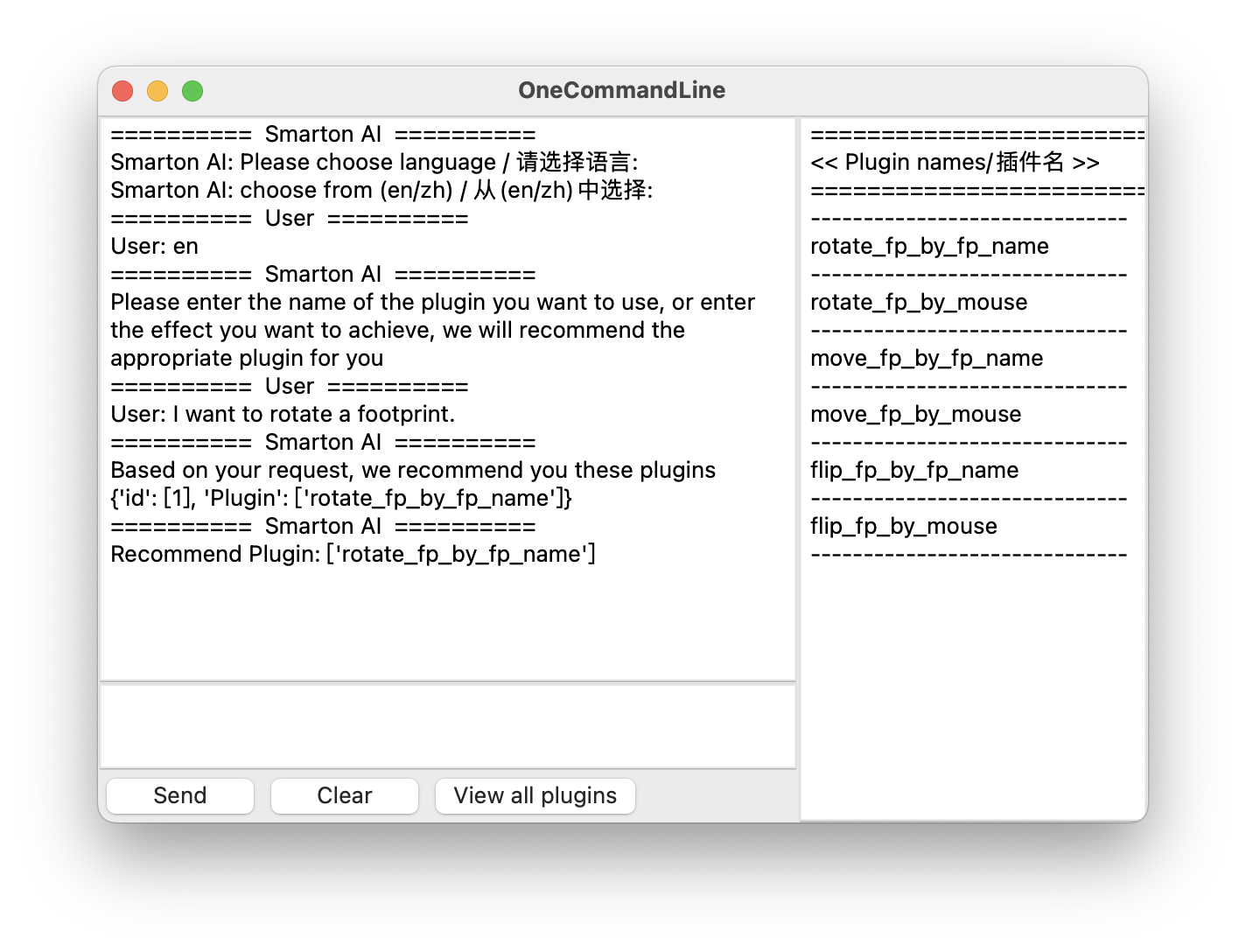}
    \caption{Plugin Matching and Selection from User Input}
    \label{fig:plugin_selection}
\end{figure}

\begin{figure}[ht]
    \centering
    \includegraphics[width=0.48\textwidth]{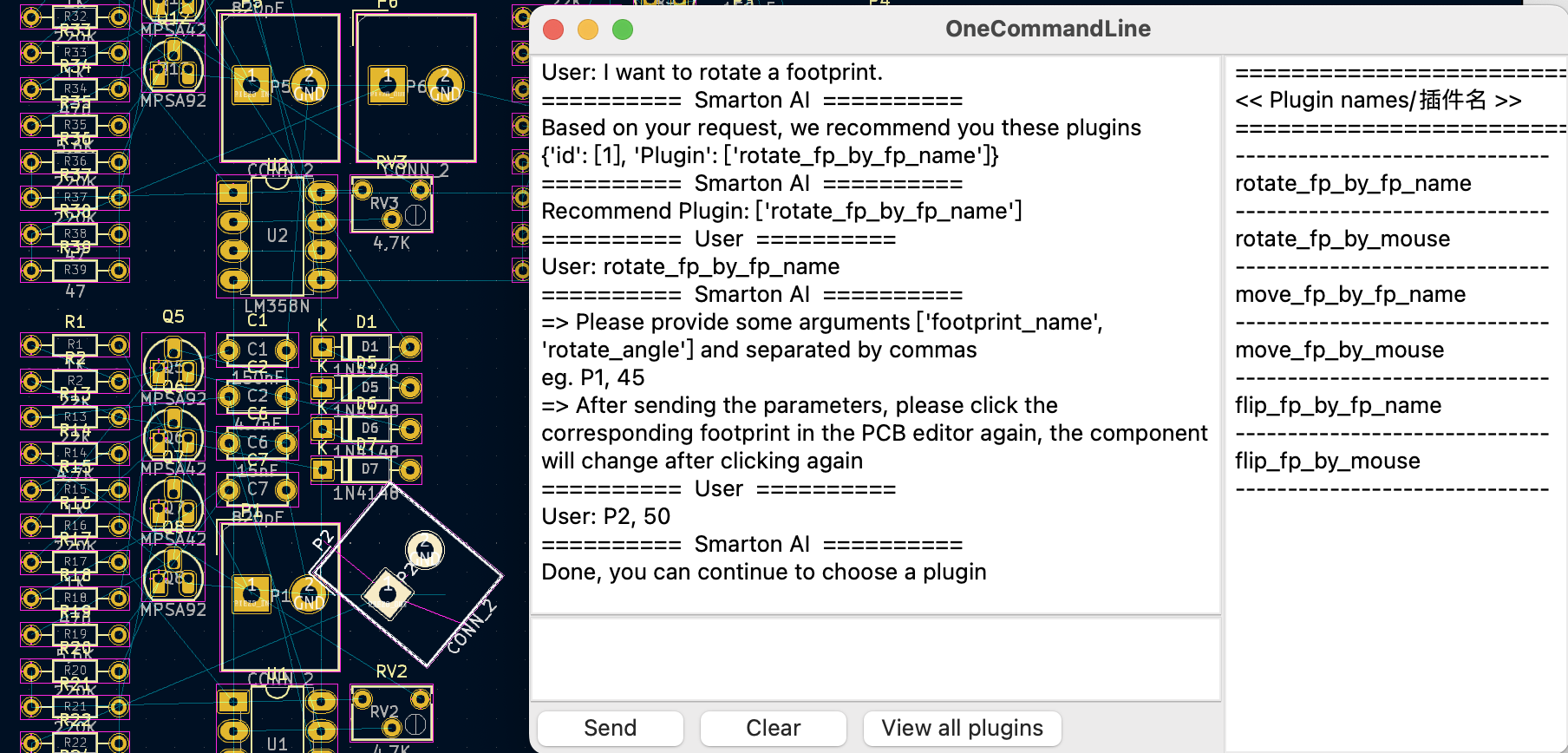}
    \caption{Plugin Execution within PCB Editor via OneCommandLine Plugin}
    \label{fig:plugin_execution}
\end{figure}

Figures~\ref{fig:plugin_selection} and~\ref{fig:plugin_execution} demonstrate SmartonAI’s OneCommandLine Plugin, designed for low-friction, single-turn plugin invocation. Given a natural language command (e.g., “rotate a footprint by 90 degrees”), the system performs intent classification, ranks available KiCad plugins using semantic similarity and metadata priors, and elicits argument fields interactively.

The final command is composed and dispatched via KiCad’s embedded Python API. Execution feedback is shown in-line, and any argument correction or re-dispatch is supported through lightweight dialogue repair strategies. The plugin recommendation module employs a combination of retrieval-augmented ranking and few-shot prompting to achieve robustness across diverse user phrasing.

\subsection{Qualitative Feedback and Usability Insights}
Qualitative logs indicate that SmartonAI maintains coherent task threads across multi-turn sessions and exhibits strong generalization to diverse EDA intents. In particular, the Chat Plugin effectively disambiguates user goals and scaffolds complex workflows with minimal guidance. Internal user feedback collected during pilot deployments highlighted reductions in search latency, improved tool discoverability, and better alignment with user mental models of design processes.

Additionally, SmartonAI's document rendering and snippet grounding significantly reduced reliance on external browsing, enabling in-context learning without cognitive context switching.

\subsection{Limitations and Future Evaluation Plans}
Despite its capabilities, SmartonAI currently focuses on KiCad-specific workflows. Scaling to other EDA tools (e.g., Altium Designer, Cadence Allegro) will require modular backend extensions and retraining of retrieval pipelines on domain-specific corpora. Furthermore, occasional hallucinations during synthesis of undocumented plugin commands indicate the need for constrained decoding or retrieval-aware decoding heads.

To further quantify system performance, we plan controlled usability studies using the System Usability Scale (SUS) and NASA-TLX metrics, along with task success rate and dialogue turn count as operational metrics. Additionally, we aim to fine-tune SmartonAI components using collected user traces under a curriculum learning regime to optimize performance across difficulty tiers.

\section{Conclusion}

In this work, we present \textbf{SmartonAI}, an intelligent assistant that augments Electronic Design Automation (EDA) workflows through retrieval-augmented generation, modular agent design, and interactive plugin execution within the KiCad ecosystem. By leveraging a hybrid LLM backend and document-grounded reasoning, SmartonAI effectively bridges the gap between natural language queries and executable design actions.

Through comprehensive experiments and qualitative feedback, we demonstrate the system’s ability to decompose complex multi-turn instructions, recommend relevant tools, and dynamically adapt its behavior to user intent. The integration of DocHelper enables fine-grained retrieval from structured documentation, reducing cognitive load and enhancing usability in practical design scenarios.

While current evaluations focus on KiCad, the underlying architecture is extensible to other EDA platforms via API adapters and domain-specific retrievers. Future work will involve fine-tuning with real user interaction traces, scaling to multimodal inputs (e.g., schematics or layout previews), and conducting large-scale user studies to quantify impact on productivity and decision support.

Our findings suggest that SmartonAI offers a promising paradigm for embedding conversational intelligence directly into professional design tools, thereby enabling more accessible, efficient, and context-aware EDA experiences.

\bibliographystyle{icml2025}
\bibliography{reference}

\end{document}